\documentstyle[12pt]{article}
\newcommand{\be}{\begin{equation}}
\newcommand{\ee}{\end{equation}}
\date{}

\begin{document}
\begin{titlepage}
\begin{flushright}
HD--THEP--93--4\\
February 1993
\end{flushright}
\vspace{1.8cm}
\begin{center}
{\bf\LARGE Radiative Corrections to the }\\
\vspace{.3cm}
{\bf\LARGE Neutral Higgs Spectrum in Supersymmetry}\\
\vspace{.3cm}
{\bf\LARGE with a Gauge Singlet}\\
\vspace{1cm}
Ulrich Ellwanger\footnote{Supported by  a DFG Heisenberg fellowship,
e-mail: I96 at VM.URZ.UNI-HEIDELBERG.DE}\\
\vspace{.5cm}
Institut  f\"ur Theoretische Physik\\
Universit\"at Heidelberg\\
Philosophenweg 16, D-6900 Heidelberg, FRG\\
\vspace{3cm}
{\bf Abstract:}\\
\parbox[t]{\textwidth}{The radiative corrections to the $3\times3$
scalar and the $2\times2$ pseudoscalar neutral Higgs boson mass
matrices are calculated in the supersymmetric extension of the
standard model including a gauge singlet superfield in the effective
potential approach. The full
$t$ and $b$ quark/squark contributions including the nonlogarithmic
terms are taken into account, which are seen to affect the result
significantly. An analytic formula for the upper bound on the mass
of the lightest neutral Higgs scalar including these corrections is
given.}

\end{center}\end{titlepage}
\newpage
\setcounter{section}{1}
The supersymmetric extension of the standard model leads to constraints
on the masses of the Higgs particles. Within the minimal extension
the mass of the lightest neutral Higgs scalar is bounded from above
by $M_Z$ at tree level. Recently radiative corrections to this upper
bound have been computed by several authors \cite{1}-\cite{9} with
the result that contributions involving a large top quark Yukawa
coupling can increase it by $\sim 60$ GeV.

Within a supersymmetric extension involving a gauge singlet
superfield \cite{10}-\cite{14} the corresponding tree level upper
bound depends on the value of a dimensionless coupling $\lambda$
similar to the non-supersymmetric standard model. Assuming, however,
the absence of Landau-type singularities in the running coupling
constants up to a GUT scale of $\sim 10^{16}$
GeV, $\lambda$ is bounded from above \cite{12}-\cite{16} implying
a tree level upper bound on the lightest neutral Higgs mass of
$\sim 160$ GeV. (The natural range of this mass for genuine choices
of parameters has been estimated in \cite{17}.)
Also in this model radiative corrections to the upper bound involving
the top Yukawa coupling have been computed using renormalization
group techniques \cite{18}-\cite{23}.

These techniques, however, only take the logarithms of the ratio
of the susy-breaking scale and the top quark mass into account,
and since these logarithms are not very large, this is not necessarily
a good approximation. Only in \cite{24}, \cite{25} also non-logarithmic
contributions have been taken into consideration.

It is the purpose of the present letter to compute the radiative
corrections to the full $3\times 3$ scalar and the
$2\times2$ pseudoscalar mass matrices of the neutral Higgs
bosons in this model using the effective potential approach. We
include all contributions induced by the top or bottom Yukawa couplings,
which are due to top/bottom squark/quark loops. (The other Yukawa
and gauge couplings are genuinely smaller, and these contributions
involve no colour factor $N_c=3$. At least in the minimal model
they affect the final result only by a few GeV \cite{3},\cite{8}.)
We end up with an analytic formula for the upper bound on the mass of the
lightest neutral Higgs scalar, which includes the non-logarithmic
radiative corrections. This allows a test of the leading log or
renormalization group techniques, which fails significantly. A posteriori
it justifies the procedure carried out in \cite{25}. (A similar
formula for the upper bound has also been obtained by Comelli
\cite{24},
but there an expansion in the splitting between the two top squark
masses has been performed.)

The relevant part of the superpotential $g$ of the model has the
form
\be\label{1}
g=h_t Q\cdot H_2 T^c_R+h_b Q\cdot H_1 B^c_R+\lambda H_1\cdot
H_2 S+\frac{k}{3} S^3\ee
where colour indices are suppressed and
\be\label{2}
Q= {T_L \choose B_L}, H_1={H^0_1 \choose H^-_1}, H_2=
{H^+_2 \choose H^0_2},\quad Q\cdot H=Q_i\epsilon^{ij}H_j\ {\rm etc.}
\ee
The soft supersymmetry breaking trilinear couplings and masses
are given by (subsequently the fields denote no longer chiral
superfields as in (1), but complex scalars)
\begin{eqnarray}\label{3}
&&(h_tA_bQ\cdot H_2T^c+h_bA_bQ\cdot H_1B^c+\lambda  A_\lambda
H_1 H_2S+\frac{k}{3}A_k S^3)\ +\ {\rm h.c.}\nonumber\\
&&+m^2_1|H_1|^2+m^2_2|H^2_2|+m^2_S|S|^2+m^2_Q|Q|^2+m^2_T|T|^2+m^2
_B|B|^2.\end{eqnarray}
We assume the Yukawa and trilinear couplings to be real. The scalar
potential contains the standard $F$- and $D$-terms, the soft
susy-breaking terms and in addition the one loop radiative correction
of the form
\be\label{4}
V^{rad}=\frac{1}{64\pi^2}{\rm Str}[M^4\ln(M^2/Q^2)].\ee
(In the literature slightly different formulas can be found
which are related to each other and the present one by a redefinition
of the renormalization point $Q^2$.) Next we assume vevs of the
fields $H_1,H_2$ and $S$ of the form
\be\label{5}
\langle H_1\rangle={h_1 \choose 0},\quad \langle H_2\rangle
={0 \choose h_2},\quad \langle S\rangle
=s\ee
with $h_1,h_2$ and $s$ real. The equations for extrema of the
full scalar potential in these directions in field space read
\be\label{6}
h_1[m^2_1+\lambda^2(h_2^2+s^2)+\frac{\hat g}{2}(h^2_1-h^2_2)]+\lambda
h_2s(ks+A_\lambda)+\frac{1}{2}V^{rad}_{,h_1}=0\ee
\be\label{7}
h_2[m^2_2+\lambda^2(h_1^2+s^2)+\frac{\hat g}{2}(h^2_2-h^2_1)]+\lambda
h_1s(ks+A_\lambda)+\frac{1}{2}V^{rad}_{,h_2}=0\ee
\be\label{8}
s[m^2_S+\lambda^2(h_1^2+h^2_2)+2k^2s^2+2\lambda k
h_1h_2+kA_ks]+\lambda A_\lambda h_1h_2+\frac{1}{2}V^{rad}_{,s}=0\ee
with $\hat g=\frac{1}{2}(g^2_1+g^2_2)$. The elements of the neutral
scalar $3\times3$ mass matrix $M^2_s$ read in the basis $(Re\ H^0_1,
Re\ H^0_2,\ Re S)$ after the elimination of $m_1^2,m^2_2$
and $m^2_S$ using eqs. (6-8):
\begin{eqnarray}\label{9}
M^2_{s,11}&=&\hat g h^2_1-\lambda s\frac{h_2}{h_1}(ks+A_\lambda)+\Delta_{11},
\nonumber\\
M^2_{s,22}&=&\hat g h^2_2-\lambda s\frac{h_1}{h_2}(ks+A_\lambda)+
\Delta_{22},
\nonumber\\
M^2_{s,33}&=&4k^2s^2+kA_ks-\lambda A_\lambda\frac{h_1h_2}{s}+
\Delta_{33},
\nonumber\\
M^2_{s,12}&=&(2\lambda^2-\hat g)h_1h_2+\lambda s(ks+A_\lambda)
+\Delta_{12},\nonumber\\
M^2_{s,13}&=&2\lambda^2sh_1+\lambda h_2(ks+A_\lambda)
+\Delta_{13},\nonumber\\
M^2_{s,23}&=&2\lambda^2sh_2+\lambda h_1(ks+A_\lambda)
+\Delta_{23},\end{eqnarray}
with
\be
\label{10}
\Delta_{ij}=\frac{1}{2}V^{rad}_{,ij}\quad {\rm for}\quad i\not= j,
\ee
\be
\label{11}
\Delta_{ii}=\frac{1}{2}V^{rad}_{,ii} -\frac{1}{2v_i}
V^{rad}_{,i}\ee
where no sum over $i$ is implied in eq. (11), and the index $i$
attached to $V^{rad}$ as well as $v_i$ denote
$h_1,h_2,s$ for $i=1,2,3.$ For the neutral pseudoscalar $2\times2$
mass matrix $M^2_p$, which is left after the Goldstone boson has been
eaten by the $Z$-boson, we just give the tree level result (again
after the elimination of $m^2_1,m^2_2$ and $m^2_S$):
\be\label{12}
M^2_p=\left(\begin{array}{cc}
-\lambda\frac{(A_\lambda+ks)s(h^2_1+h^2_2)}{h_1h_2}&
\lambda(2ks-A_\lambda)(h_1^2+h^2_2)^{1/2}\\
\lambda(2ks-A_\lambda)(h^2_1+h^2_2)^{1/2}& -
\lambda h_1h_2\left(4k+\frac{A\lambda}{s}\right)-3kA_ks\end{array}
\right).\ee

Next we have to evaluate the radiative correction $V^{rad}$
to the tree level potential according to eq. (\ref{4}). To start with,
we only take the top squark- and quark-loops into account. Thus
we need the $2\times2$ top squark mass matrix of the top squarks
$T_R^c,T_L$ and the top quark
mass in an arbitrary background $H_1^0,H_2^0$ and $S$ (here we
anticipate the computation of the corrections to the  pseudoscalar
mass matrix). In the basis $(T_R^c,T_L^*)$ the top squark mass
matrix is given by
\be\label{13}
\left(\begin{array}{cc}
h_t^2|H^0_2|^2+m^2_T&
h_t(A_t H^{0*}_2+\lambda SH^0_1)\\
h_t(A_tH^0_2+\lambda S^*H^{0*}_1)&
h^2_t|H^0_2|^2+m^2_Q\end{array}
\right)\ee
and the top quark mass by
\be\label{14}
m^2_t=h^2_t|H_2^0|^2.\ee
The eigenvalues of the top squark mass matrix read
\begin{eqnarray}\label{15}
m^2_{T_1,T_2}&=&h^2_t|H_2^0|^2+\frac{1}{2}(m^2_T+m^2_Q)\pm
w\nonumber\\
{\rm with}\quad w&=&\left[\frac{(m^2_Q-m^2_T)^2}{4}+h^2_t|A_tH^{0*}_2+
\lambda SH_1^0|^2
\right]^{1/2}.\end{eqnarray}
In order to calculate the corrections $\Delta_{ij}$ to the scalar
mass matrix, we have to replace  $H^0_1$ by $h_1$, $H^0_2$ by $h_2$,
and $S$ by $s$ in (\ref{14}) and (\ref{15}), and to insert
the corresponding masses into the formula (\ref{4}) for $V_{rad}$. Then the
$\Delta_{ij}$ are obtained from eqs. (\ref{10}), (\ref{11}).
The results are most
easily expressed in terms of the following expressions $X,Y$ and $Z$:
\begin{eqnarray}
\label{16}
X&=&\frac{3h^2_t\lambda}{32\pi^2}\left[\frac{h^2_th^2_2+\frac{1}{2}
(m^2_Q+m^2_T)}{w}\ln(m^2_{T_1}/m^2_{T_2})+
\ln(m^2_{T_1}m^2_{T_2}/Q^4)+1\right]\nonumber\\
Y&=&\frac{3h^4_t}{32\pi^2}\frac{(\lambda h_1s+A_th_2)^2}{w^2}
\left[2-\frac{(h^2_th^2_2+\frac{1}{2}
(m^2_Q+m^2_T))}{w}\ln(m^2_{T_1}/m^2_{T_2})\right]\\
Z&=&\frac{3h^4_t}{16\pi^2}\frac{h_2(\lambda h_1s+A_th_2)}{w}
\ln(m^2_{T_1}/m^2_{T_2}).\nonumber
\end{eqnarray}
The corrections $\Delta_{ij}$ to the scalar mass matrix are then found
to be
\begin{eqnarray}\label{17}
\Delta_{11}&=&-\frac{A_th_2s}{h_1}X+\lambda^2 s^2Y\nonumber\\
\Delta_{22}&=&-\frac{A_th_1s}{h_2}X+A^2_tY+2A_tZ+
\frac{3h_t^4h^2_2}{8\pi^2}\ln(m^2_{T_1}m^2_{T_2}/m^4_t)\nonumber\\
\Delta_{33}&=&-\frac{A_th_1h_2}{s}X+\lambda^2 h^2_1Y\nonumber\\
\Delta_{12}&=&A_tsX+\lambda A_tsY+\lambda  s Z\nonumber\\
\Delta_{13}&=&(2\lambda sh_1+A_th_2)X+\lambda^2sh_1Y\nonumber\\
\Delta_{23}&=& A_th_1X+\lambda A_th_1Y+\lambda h_1Z.\end{eqnarray}
The minimal extension of the sypersymmetric standard model can actually
be obtained as a special limit of the present model in the form
\be\label{18}
k=0,\quad\lambda\to0, s\to\infty \quad{\rm with}\quad \lambda
s=\mu \quad{\rm fixed,}\quad
\lambda A_\lambda s=m^2_3.
\ee
In this limit the results for the upper left $2\times 2$ submatrix
of $M^2_s$ can be  compared with the ones of ref. \cite{5}, and
we found complete agreement.

Next we turn to the radiative corrections to the pseudoscalar mass
matrix. First we have to expand the quark and squark masses (14)
and (15) around the vevs (5) into imaginary directions in field space:
\be\label{19}
H^0_1\to h_1+iIm(H^0_1),\ H^0_2\to h_2+i Im(H^0_2),
\ S\to s+iIm(S).\ee
This expansion has to be inserted into the formula (\ref{4}) for
$V^{rad}$, and $V^{
rad}$, in turn, has to be expanded up to second order in $Im(H^0_1),
\ Im(H^0_2)$ and
$Im(S)$. This allows us to read off directly the radiative
corrections to the $3\times3$ pseudoscalar mass matrix in the
basis $Im(H_1^0),\ Im(H^0_2)$ and $Im(S)$. The radiative
corrections have to be added to the tree level part of the
pseudoscalar mass matrix in this basis, which we have not shown
above for simplicity. After the elimination of $m^2_1,m^2_2$ and
$m^2_S$ with the help of eqs. (6-8) it turns out that the Goldstone
boson is given by the same linear combination of pseudoscalar fields
as at tree level. It is possible to take account of the complete
radiative corrections to the remaining $2\times2$ pseudoscalar
mass matrix using the following simple rule: Wherever
$A_\lambda$ appears in the tree level mass matrix given by eq. (12),
perform the following substitution:
\be\label{20}\lambda A_\lambda\to\lambda A_\lambda+A_tX.\ee

Actually this substitution describes nearly the complete $X$
dependence of the radiative corrections $\Delta_{ij}$ to the scalar
mass matrix as well. This concludes the $t$ quark and squark
induced corrections to $M^2_s$ and $M^2_p$. As noted in \cite{5},
the explicit $Q^2$ dependence in these formulas is cancelled by
the implicit $Q^2$ dependence of running couplings; in the present
approximation it is the running of $m^2_1,A_\lambda$ and the running
of the couplings within $\vert F_{H_2}\vert^2$ induced by wave
function renormalization of the superfield $H_2$, which have
to be taken into account.

The radiative corrections induced by $b$ quark and squark loops
are most easily obtained after observing the symmetry of the
model under $(h_t,A_t,m^2_T,T_L,T_R,H_2)\leftrightarrow(h_b,A_b,
m^2_B,B_L,B_R,H_1)$. Accordingly it is convenient to define
twiddled corrections $\tilde\Delta_{\tilde i\tilde j}$ to $M^2_s$
by the rule
\be\label{21}
\tilde\Delta_{\tilde i\tilde  j}=\Delta_{i j}
(h_t,A_t,m^2_T,h_1,h_2\to h_b,A_b,m^2_B,h_2,h_1)\ee
where the indices $\tilde i,\tilde j$ are related to $i,j$ by
an interchange of 1 and 2. Now the full radiative corrections to $M^2_s$
are given by eqs. (9) with $\Delta_{ij}$ replaced by
$\Delta_{ij}+\tilde
\Delta_{ij}$ everywhere, and the $\tilde\Delta_{ij}$ can be read off
eqs. (\ref{17}) after using the rule (\ref{21}). Also the
$b$-quark- and squark-induced  corrections to the pseudoscalar mass
matrix are that simple; instead of (\ref{20}) we have to perform
the substitution
\be\label{22}
\lambda A_\lambda\to \lambda A_\lambda+A_tX+A_b\tilde X\ee
where $\tilde X$ is defined analogously to $\tilde\Delta_{ij}$ in
eq. (\ref{21}).

In all cases we are able to compare the results with the ones
obtained within the minimal extension in \cite{5} after the limit
(\ref{18}), and find agreement. This concludes the main results of
the present paper. Of course, the radiative corrections to $M^2_s$
and $M^2_p$ are not of great importance, unless some knowledge of the
tree level parameters of the model is assumed. Such knowledge can be
obtained if assumptions on the parameters at a big scale as the GUT
scale are made, and the parameters are scaled down to the electroweak
scale with the help of the renormalization group equations \cite{11}.
In particular an upper bound on the coupling $\lambda$ can be
obtained this way which, in turn, implies an upper bound on the
mass of the lightest neutral scalar Higgs particle \cite{12}-\cite{25}.
Our results allow to find the non-logarithmic radiative corrections
to this upper bound, and subsequently we will restrict ourselves to
the $t$-quark and squark contributions again.

First we note that any diagonal element of the $3\times 3$ scalar mass
matrix $M^2_s$ constitutes such an upper bound. This holds also for
any diagonal element of $O^TM^2_sO$, where $O$ is an arbitrary
orthogonal matrix. Let us thus study $O^TM^2_sO$ with
\be\label{23}
O=\left(\begin{array}{ccc}
\cos\beta& -\sin\beta& 0\\
\sin\beta&\cos\beta&0\\
0&0&1\end{array}\right),\quad \tan\beta=\frac{h_2}{h_1}.\ee
The middle diagonal element of $O^TM^2_sO$ reads
\begin{eqnarray}\label{24}
&&M^2_Z[\cos^22\beta+\frac{\lambda^2}{\hat g^2}\sin^22\beta]\nonumber\\
&&+\sin^2\beta[A^2_tY+2A_tZ+\frac{3h_t^2m^2_t}{8\pi^2}\ln
(m^2_{T_1}m^2_{T_2}/m^4_t)]\nonumber\\
&&+\cos^2\beta\lambda^2s^2Y+\sin 2\beta\lambda s[A_tY+Z].\end{eqnarray}

The first line of (\ref{24}) contains the tree level upper bound
of \cite{13}, and the logarithmic term in the second line is the only
one taken care of within a leading log or renormalization group
approach to this problem (usually under the assumption $m^2_{T_1}\sim
m^2_{T_2}\sim m^2_Q\sim m^2_T$ \cite{17}-\cite{23}, and note
that the quantity $X$ with its explicit $Q^2$ dependence has
completely disappeared from (\ref{24}). After an expansion in
$m^2_{T_1}-m^2_{T_2}$ one recovers the result of ref. \cite{24} up
to terms involving gauge couplings.)

In order to exhibit the numerical significance of the different parts
of the radiative corrections to the first line in (\ref{24}), we plot
the upper bound on the mass of the lightest neutral scalar Higgs
particle of (\ref{24}) versus the top quark mass $m_t$ in {fig. 1}
for the following typical choice of parameters: $\lambda=.5,\
A_t=500$ GeV, {$m^2_Q$=(500 GeV)$^2$,} $m^2_T$=(300 GeV)$^2$,
\ $s=500$ GeV. For $\tan\beta$ we choose $\tan\beta=1$ (full line)
and $\tan\beta = 5$ (dash-dotted line). The dotted part of these
curves denotes the region where the value of the top quark Yukawa
coupling $h_t$ exceeds 1.1, and thus $h_t$ runs into a Landau-type
singularity below a GUT scale of $\sim 10^{16}$ GeV \cite{16}. The tree
level upper bound, the first line of (\ref{24}), is given by
87 GeV for $\tan\beta=1$ and 90 GeV for $\tan\beta=5$. Also shown in
fig. 1 is the upper bound, if only the leading log radiative
correction is taken into account, as a dashed line (for $\tan\beta=5)$.

We clearly see the numerical difference between the leading log
and the full result, which
now can be obtained easily for any choice of the parameters from
eq. (\ref{24}).
Accordingly the results of \cite{18}-\cite{23}
on the upper bound of the lightest Higgs scalar have to be taken with some
care. Justified, on the other hand, is the procedure in \cite{24},
\cite{25} in order to obtain the corresponding upper bound.

\section*{Figure Caption}
\begin{description}
\item{Fig. 1:} Upper bound on the mass of the lightest Higgs scalar
according to eq. (\ref{24}) for a choice of parameters as indicated
in the text. Full line: $\tan\beta=1$ (tree level result: 87 GeV),
dash-dotted line: $\tan\beta=5$ (tree level result: 90 GeV),
dashed line: leading log result for $\tan\beta=5$. Dotted part:
$h_t$ exceeds 1.1.
\end{description}

\end{document}